# AN EXPLORATION OF ROTATING LEADERSHIP IN A KNOWLEDGE BUILDING COMMUNITY

Leanne Ma
Yoshiaki Matsuzawa
Derya Kici
Marlene Scardamalia
University of Toronto/Ontario Institute for Studies in Education
252 Bloor Street West
Toronto, Ontario, M5S 1V6, Canada
E-mail: leanne.ma@mail.utoronto.ca

## ABSTRACT
This study aims to investigate the COINs concept of rotating leadership within a Knowledge Building context. Individual and group level leadership patterns in a grade 4 science class were explored through temporal visualization of betweenness centrality. Results indicate that the student network was relatively decentralized, with almost all students leading the group at different points in time. Rotating leadership appears to be an emergent phenomenon of Knowledge Building, and we suggest that it has the potential to be an indicator of collective cognitive responsibility.

## INTRODUCTION
Knowledge Building (KB; Scardamalia, 2002) is a principle-based pedagogy that engages students directly in processes of knowledge creation through sustained, creative work with ideas. Within a KB community, students assume collective cognitive responsibility for the advancement of their community knowledge. Thus, the success of the community lies in the distribution of group effort across all members, rather than a concentration of effort amongst a few individuals. For example, a case study analysis of a KB class revealed that the decentralization of the student network led to increased advances in collective knowledge (Zhang, Scardamalia, Reeve, & Messina, 2009). This observation is in line with research on innovative groups in open network communities, as research in COINs theory (Gloor, Laubacher, Dynes, & Zhao, 2003) shows that the decentralization of leadership across group members through patterns of oscillation is a good indicator of group productivity and creativity. The current exploratory study aims to use COINs theory as an analytical framework to investigate leadership dynamics in KB. We used social network analysis (SNA) in order to visualize patterns of collective cognitive responsibility over time. We expected that rotating leadership, as indicated by oscillating betweenness centrality similar to those of productive innovative groups (Kidane & Gloor, 2007), would emerge as the collective knowledge advances.

## METHODS AND ANALYSIS
Our sample is a grade 4 class, comprising 22 students, who engaged in the inquiry of light over a three-month period (Sun, Zhang, & Scardamalia, 2010). Student discourse in an online KB environment (i.e., Knowledge Forum®) was exported into Knowledge Building Discourse Explorer (KBDeX; Oshima, Oshima, & Matsuzawa, 2012) in order to perform SNA based on a list of content-related words (101 words) compiled from the Ontario Curriculum of Science and Technology. KBDeX is an analytic tool which was designed to facilitate content-based SNA for KB discourse. It produces visualization models of word, note, and student networks based on the co-occurrence of words in a note. Edges in the student network show the strength of connections among students whose notes share the same word. KBDeX also supports temporal network analysis by showing animations of how the network metrics (e.g., betweenness centrality) change over time. The betweenness centrality measures the extent to which a member influences other members of the group (Gloor et. al., 2003). At the individual level, a betweenness centrality value of 1 means that a member is highly influential, whereas a value of 0 means that a member is equally influential as other members. At the group level, the centralization of betweenness centrality is used to indicate the extent to which the network is centralized (i.e., influence is not evenly distributed in the network, high influence is occupied by a few members of the network). A centralization of betweenness centrality value of 1 means that the network is completely centralized, whereas a value of 0 means that the network is completely decentralized.

## FINDINGS AND FUTURE DIRECTIONS
First, we examined the centralization of the betweenness centrality of the student network in order to examine the extent to which the student network was centralized around specific members.

We found that the average centralization of betweenness centrality is relatively low, $m = 0.089$, $sd = 0.064$ (range = 0 to 0.25), which suggests that the student network was relatively decentralized, and students shared more or less the same level of influence in their network. Students maintained a relatively high cohesive network over time.

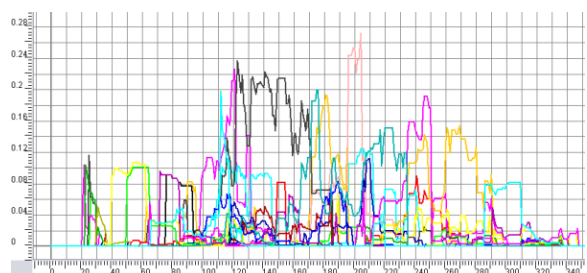

*Figure 1. KBDeX visualization of individual betweenness centralities across time*

Next, we examined the betweenness centrality at the individual level as shown in Figure 1. The Y axis of the chart shows the betweenness centrality value, and the X axis shows the time. Each student is represented by a coloured line, resulting in a total of 22 lines displayed in the chart. Of the 22 students, 20 students took a leading position, suggesting that different students were leading across different times. As a higher betweenness centrality means that the student holds a more important position in the social network, the overlapping phenomena of the lines indicate that the leading student (i.e., the student with the highest centrality at the time) is frequently changing.

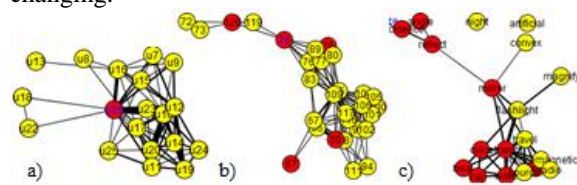

*Figure 2: a) student network, b) note network, and c) word network at turn 112, when student u26 had the highest betweeness centrality*

Follow up content analyses on the top five cases of betweenness centrality (range = 0.18 to 0.25) revealed that students with the highest betweenness centrality held a pivotal role in connecting diverse student ideas with the whole-class discussion. The most typical case was selected for explanation and the three networks in KBDeX are shown in Figure 2. The student network in Figure 2a shows that student u26 connected students u18 and u22 to the largest group network. The note network in Figure 2b shows that notes 78 and 120, written by student u26, linked notes 72, 73, and 119 to the larger cluster of notes.

The word network in Figure 2c shows that student u26 connected the concepts of "reflection", "angle", and "mirror" to the main discussion of whether light travels in a wave or straight line. The discussion later touched upon topics such as "mirrors and angles", "reflection and absorption", and "colours and shadows". Thus, student u26 helped the entire class reach a deeper understanding of light, by seeking coherence between diverse ideas from student u18, student u22, and the whole group. Similar patterns were found with the other four cases of leadership.

Our results confirm our hypothesis that rotating leadership is manifested within a KB community that values idea diversity and assumes collective cognitive responsibility for idea improvement. Future research should further examine the conditions that give rise to the phenomenon of rotating leadership. For example, it would be interesting to supplement our findings with additional qualitative analyses in order to detangle whether or not other types of contribution lead to rising or falling leadership amongst group members.